\documentstyle[prb,aps]{revtex}   
\begin{document}   
   
\draft   
\title{A Review of the Real Space Renormalization Group Analysis of the Two-
Dimensional Coulomb Gas, the Kosterlitz-Thouless-Berezinski Transition, and 
Extensions to a Layered Vortex Gas}   
   
\author{Stephen W.~Pierson\cite{email}}   
\address{Department of Physics, Worcester Polytechnic Institute, Worcester, 
MA 01609-2280}   
  
\date{\today}   
\maketitle   
\begin{abstract} 
Fundamental properties of the Kosterlitz-Thouless-Berezinskii (KTB) transition 
which occurs in systems in the universality class of the two-dimensional $X$-$Y$ 
model are reviewed here with an emphasis on the real-space renormalization 
group analysis used to derive the recursion relations. A derivation of the unique 
temperature dependence of the correlation length of this system will be presented 
and the signatures of the KTB transition in electronic transport measurements 
will be explained. Extensions of this model and technique to layered 
superconductors will also be presented. The size of the 3D critical region, the 
nature of the 3D crossover, and the effect of a finite electric current will be 
discussed. 

\end{abstract}
   
\section{Introduction}   
\label{sec:intro}   
The phase transition of the systems in the universality class of the two-
dimensional (2D) $X$-$Y$ model, known as the Kosterlitz-Thouless-Berezinskii 
(or some permutation of this) transition (Berezinskii 1971; Kosterlitz and 
Thouless 1973; Kosterlitz 1974), is a fascinating one. And, even though the basic 
details of this transition were worked out in the early 1970's, interest in this 
transition remains keen with more than 200 articles published yearly on the topic. 
Indeed, one can find references to the ``KTB" transition in the literature on topics 
ranging from layered materials (such as the high-temperature superconductors 
(HTSC's)) to superfluid films. Given its wide-spread importance, an 
understanding of the Kosterlitz-Thouless-Berezinskii transition and the derivation 
of its critical behavior properties is essential for a student of condensed matter 
and statistical physics. Renormalization group (RG) techniques (Wilson 1971) are 
also an important tool for a theoretical physicist and can be used for a variety of 
problems in a number of ways. [For example, see the article by N. Andrei in this 
same issue.] It was originally applied to the 2D $X$-$Y$ model by Kosterlitz 
(1974) to study its critical behavior.  In this paper, we will review his real-space 
RG analysis of the 2D $X$-$Y$ model, describe the quintessential properties of 
the critical behavior, and discuss recent results on layered superconductors 
derived using the same mathematical formulation.

\section{The 2D $X$-$Y$ Model and the Real Space Renormalization Group Analysis}   
\label{sec:model}
Falling into the universality class of the 2D $X$-$Y$ model are 2D 
superconductors and superfluids as well as 2D two-component magnetic systems. 
More generally, one can say that this universality class covers 2D systems with a 
two-component order parameter. Systems in this universality class are believed to 
undergo the Kosterlitz-Thouless-Berezinskii phase transition. In this section, we 
will show how the 2D $X$-$Y$ model can be mapped to the 2D Coulomb gas 
and then review Kosterlitz's RG study of that system which illuminates many of 
the principal features of the critical behavior. (For pedagogical arguments of the 
phase transition see Kosterlitz et al., 1973 or Halperin (1979).) The RG study 
culminates in the 2D recursion relations which can be used to derive the 
temperature dependence of the correlation length. In the last subsection we will 
describe how KTB behavior can be detected in superconducting films via 
electronic transport measurements.

One might guess that the physics of the 2D $X$-$Y$ model is uninteresting since 
it has been shown that there is no long-range order at finite temperature in 
magnetic systems (Mermin and Wagner 1966) nor in superfluids (Hohenberg 
1967). Yet hints of novel behavior were provided by Stanley and Kaplan (Stanley 
and Kaplan 1966) who found a divergence of the susceptibility based on a high-
temperature series expansion. Wegner (Wegner 1967) went on to show that the 
susceptibility was infinite at low temperatures. Further suggestions of a phase 
transition were provided by Berezinskii (Berezinskii 1971) who found a power-
law decay of the spatial correlation function at low temperatures while others had 
ascertained the exponential decay at high temperatures. 

It was Kosterlitz and Thouless (Kosterlitz and Thouless 1973) who formulated the 
transition explicitly in terms of vortices and, to be more precise, as a vortex/anti-
vortex pair unbinding transition.  In this zero-field transition, vortices are created 
spontaneously in pairs at finite temperatures and have an energy which goes as 
the logarithm of their separation. As the temperature increases, the vortex pairs 
increase in size and in density. As this happens, screening of their interactions 
becomes important and enables a phase transition at a temperature $T_{KTB}$. 
At this temperature two things happen: vortex pairs start to unbind and 
spontaneous creation of single free vortices becomes possible. Being driven by 
topological excitations is one of the properties of this transition that makes it so 
unique. Another is its correlation length which diverges according to an 
exponential rather than a power-law at $T_{KTB}$. As we will explain, it is this 
latter property that makes detection of this phase transition difficult to detect by 
thermodynamic measurements because there is no singularity in the free energy. 
Evidence for the transition can be accumulated however through electronic 
transport measurements for the case of superconductors.

\subsection{Mapping the 2D $X$-$Y$ Model to the 2D Coulomb Gas}

The 2D $X$-$Y$ model is defined by the Hamiltonian:
\begin{equation}
H=-J\sum_{<i,j>} {\bf s}_i\cdot{\bf s}_j=-Js^2\sum_{<i,j>} cos (\phi_i-\phi_j),
\label{2dxy}
\end{equation}
where $<i,j>$ are all nearest-neighbor spins and $\phi_i$ is the angle that the 
spin vector ${\bf s}_i$ makes with an arbitrary axis. In the low temperature and 
continuous limit, one has
\begin{equation}
H=E_0+{Js^2\over 2}\int d^2r|\nabla\phi|^2,
\label{2dcontxy}
\end{equation}
where $E_0$ is a constant. One can immediately see from Eq.~(\ref{2dcontxy}) 
the similarity to a Ginzburg-Landau free energy functional in which the 
amplitude variations are neglected. We now proceed to express 
Eq.~(\ref{2dcontxy}) in terms of the energy for a 2D Coulomb gas. This is 
accomplished (Halperin 1979) by first splitting the angle $\phi$ into two parts, 
$\phi=\phi_v+\phi_{sw}$, corresponding to a {\it v}ortex part and a {\it s}pin 
{\it w}ave part where
\begin{equation}
\nabla\times\nabla\phi_{sw}=0,
\end{equation}
\begin{equation}
\nabla\cdot\nabla\phi_{v}=0.
\end{equation}
This substitution leads to three terms in Eq.~(\ref{2dcontxy}) including a cross 
term ($\int d^2r\nabla\phi_{sw}\cdot\nabla\phi_{v}$) which can be shown to be 
equal to zero after an integration by parts leaving the vortex and spin wave 
contributions independent of one another:
\begin{equation}
H=E_0+{Js^2\over 2}\int d^2r|\nabla\phi_v|^2+{Js^2\over 2}\int d^2r|\nabla\phi_{sw}|^2.
\label{2ind}
\end{equation}
It is the vortex term that interests us because, as we will show, it contains the part 
that drives the phase transition. 

The next step in getting to the 2D Coulomb gas is to assume that vortices are 
present in the system which means that the line integral of the change in the field 
$\phi_v$ around any contour $c$ is $2\pi N$ where $N$ is any integer:
\begin{equation}
\oint_c \nabla\phi\cdot d{\bf r}=2\pi N.
\label{vorticity}
\end{equation}
The simplest, non-trivial case is when the contour encloses exactly one vortex of 
vorticity one in which case $N=1$. The right hand side (RHS) of 
Eq.~(\ref{vorticity}) can be written in terms  of a vortex density $\sigma(r)$ and 
the left hand side (LHS) can be rewritten using Stoke's theorem yielding
\begin{equation}
\hat{\bf z}\cdot(\nabla\times\nabla\phi_v)=2\pi\sigma(r).
\label{stokes}
\end{equation}
By using vector identities and defining $\nabla\phi_v'=\nabla\times\nabla\phi_v$, 
Eq.~(\ref{stokes}) becomes Poisson's equation:
\begin{equation}
\nabla^2\phi_v'=2\pi\sigma(r).
\label{poisson}
\end{equation}
Using a Green's function method to solve Eq.~(\ref{poisson}), one arrives at
\begin{equation}
\phi_v'(r_1)=\int d^2r_2 \sigma(r_2)\ln|{\bf r}_1-{\bf r}_2|/\xi_0
\label{green}
\end{equation}
where $\xi_0$ is an infrared cutoff associated with the size of the core of the 
vortex. Finally, since 
$\nabla\phi_v\cdot\nabla\phi_v=\nabla\phi_v'\cdot\nabla\phi_v'$ and 
$\nabla\cdot(\phi_v'\nabla\phi_v')=\nabla\phi_v'\cdot\nabla\phi_v'+\phi_v'\nabla^
2\phi_v'$, one gets the energy for a configuration of vortices to be
\begin{eqnarray}
H=E_0+{Js^2\over 2}\int &&d^2r|\nabla\phi_{sw}|^2+2\pi{Js^2\over 2}\int d^2r\int d^2r'\sigma(r)\sigma(r')\ln|{\bf r}-{\bf r}'|/\xi_0\nonumber \\
&&+2\pi{Js^2\over 2}\int d^2r\sigma(r) \int d^2r'\sigma(r')\ln(L/\xi_0),
\label{total}
\end{eqnarray}
where $L$ is the size of the system. The first terms in Eq.~(\ref{total}) contain 
the spin wave terms and the energy associated with the vortex core energies. The 
third term is an interaction term between the vortices and the last term which 
diverges as the size of the system ensures system neutrality. 

We can now neglect the spin-wave term, assume vortex neutrality, and write the 
partition function for the 2D Coulomb gas,
\begin{eqnarray}
Z=\sum_N y^{2N}{1\over (N!)^2}
&&\int_{D_{1}} d^2r_1\int_{D_2} d^2r_2...\int_{D_{2N}} d^2r_{2N}\nonumber \\
&&\times \exp{\Bigr [} -{\beta\over 2}\sum_{i\not= j}p_i p_j V(\mid 
{\bf r}_i-{\bf r}_j \mid){\Bigr ]} 
\label{gpf}
\end{eqnarray}
where $2N$ is the total number of particles, $N$ of
which have a positive (negative) charge $p_i=+p$ ($p_i=-p$), and ${\bf
r}_i$ are the coordinates of the $i$th charge. $\beta^{-1}=k_B T$ where
$T$ is the temperature and $k_B$ is the Boltzman constant. 
$y=\exp(\beta\mu)/\tau^2$ is the fugacity, where
$\mu=- E_c$ and $E_c$ is the ``core energy,'' that is, the energy needed to 
create a normal core in a superconducting fluid (in our case).
$V(R)=-\ln(R/\tau)$ is expressed in the units $p^2$. Note that the notation ${\bf 
r}_{ij}={\bf r}_i-{\bf r}_j$ will be used.  The integrals 
are over an area $D_i$ which is all of the 
area except for disks $d_i(j)$ of radius $\tau$ around the charges 
$j<i$. The area $D_i$ is written, where $A$ is 
the total area, 
\begin{equation}
D_i=A-\sum_{j<i} d_i(j). 
\label{dii}
\end{equation}

\subsection{RG Study of the Neutral 2D Coulomb Gas}

The renormalization group technique (Wilson 1971) is useful for studying second 
order phase transitions which are scale invariant at the transition temperature.  
RG incorporates the essential idea of scaling and extends it. Scaling says that 
near a second order phase transition, one length scale, the correlation length, 
dominates the critical behavior and all other length scales are thrown out. RG 
does not eliminate the short length scales altogether as scaling does, but 
incorporates their contribution into the renormalization of the system parameters. 
There are two steps in an RG analysis: the first is to course-grain the system 
which amounts to taking out the small-scale structure and incorporating it into a 
renormalization of the parameters of the system. This can be done both in real 
space and in momentum space. (Wilson and Kogut 1974). The second step is to 
rescale the system so that the infrared cutoff is restored to its original value. The 
connection to the physics of the system is achieved by studying how the 
parameters of the system change as one goes through the RG iterations. The 
relationship between the old values and the ``renormalized'' values are called 
recursion relations. For a thorough description of the renormalization group, the 
reader should consult any of the many textbooks on the subject (e.g. Binney, 
Dowrick, Fisher, and Newman 1995; or Ma 1976). Below we sketch the details of 
the real space renormalization group study of the 2D Coulomb gas (Kosterlitz 
1974). A manuscript containing the complete details of this RG analysis and the 
derivation of the recursion relations is available from the author. The reader is 
also encouraged to see any of the excellent reviews of the KTB transition 
(Halperin 1979; Minnhagen 1987; Suzuki 1979).

The first step in the RG calculation is to integrate out small scale 
structure. In our case, this amounts to increasing the minimum separation
between the vortices $\tau$ to $\tau+d\tau$ and determining the effect of the 
vortex pairs with a separation in that range on the interactions 
of other vortices. This is realized by first increasing slightly the allowed 
size of the disks around each vortex $D_i$ in Eq.~(\ref{dii}) from $\tau$ to 
$\tau+d\tau$:
\begin{equation}
\int_{D_{i}} d^2r_i=\int_{D'_{i}} d^2r_i+\sum_{n<i}\int_{\delta_{n}(i)} 
d^2r_i.
\end{equation}
When substituted into the integrals of Eq.~(\ref{gpf}), one gets, 
\begin{eqnarray}
\int_{D_{1}}&&d^2r_1...\int_{D_{2N}}d^2r_{2N}=
\int_{D_{1}'}d^2r_1...\int_{D_{2N}'}d^2r_{2N}+{1\over 2}\sum_{i\not=j}
\int_{D_{1}'} d^2r_1 ...\int_{D_{i-1}'} d^2r_{i-1} 
\label{aab} \\
&&\times\int_{D_{i+1}'} d^2r_{i+1}...
\int_{D_{j-1}'} d^2r_{j-1} \int_{D_{j+1}'} d^2r_{j+1} ...
\int_{D_{2N}'} d^2r_{2N} \int_{D_{j}''} d^2r_j 
\int_{\delta_i({j})} d^2r_{i}+O(d\tau^2).\nonumber 
\end{eqnarray}

It is the last two integrals of the second term of the right hand side of 
Eq.~(\ref{aab}) which concern us. By doing these integrals, the effect of the small 
pairs is integrated out.  Physically, the integral over $\delta_i(j)$ puts the charge 
$i$ in an annulus around charge $j$ to form a pair of smallest separation. The 
integral over $D_j''$ then move this pair through all possible positions in the 
system. To do the integrals over ${\bf r}_i$ and ${\bf r}_j$, all of the  terms in 
the partition function that include $i$ and $j$ are isolated:
\begin{equation}
\int_{D_{j}''} d^2r_j \int_{\delta_i(j)} d^2r_{i}
\exp\Bigl[-\beta\sum_{k}p_j p_k V(r_{jk}) -\beta\sum_{k}p_i p_k 
V(r_{ik}) {\Bigr]}.
\label{aaa}
\end{equation}
One can then assume that only vortices of opposite charge can form pairs so that 
$p_i=-p_j$, 
\begin{equation}
\int_{D_{j}''} d^2r_j \int_{\delta_i(j)} d^2r_{i}
\exp\Bigl[ -\beta p_j\sum_{k} p_k [-\ln(r_{jk}/\tau)+\ln(r_{ik}/\tau)]{\Bigr]}.
\label{aba}
\end{equation}
Integrating over $r_i$, one gets ${\bf r}_i={\bf r}_j+{{\vec \tau}}$ and, 
\begin{equation}
\tau d\tau\int_{D_{j}''} d^2r_j \int_0^{2\pi}d\theta
\exp\Bigl[ -\beta p_j\sum_{k} p_k{1\over 2}\ln[(r_{jk}^2+2{\bf r}_{jk}\cdot{\vec \tau}+\tau^2) /r_{jk}^2]{\Bigr]}.
\label{aca}
\end{equation}
One then makes the approximation that the density of vortices is small and 
consequently that $r_{jk}$ is large meaning that the probability that another 
vortex is close to the pair $i$-$j$ is small. An expansion in $\tau/r_{jk}$ can 
then be performed,
\begin{equation}
2\pi\tau d\tau \int_{D_{j}''} d^2r_j 
{\biggl(}1+\int_0^{2\pi}{d\theta\over 2\pi}{(\beta p)^2\over 2}\sum_{k\not=l}p_k p_l
{\Bigl [}{ \tau r_{jk}\cos\theta\over r_{jk}^2}{\tau r_{jl}cos(\phi-\theta)\over r_{jl}^2}
- {\tau^2\cos^2\theta\over r_{jk}^2 }{\Bigr ]}{\biggr)}.
\label{cec}
\end{equation}
After the straightforward integration over $\theta$, one has,
\begin{equation}
2\pi\tau d\tau \int_{D_{j}''} d^2r_j 
{\biggl(}1+{(\beta p\tau)^2\over 4}\sum_{k\not=l}p_k p_l{\Bigl [}
{{\bf r}_{jk}\cdot{\bf r}_{jl}\over r_{jk}^2r_{jl}^2}- {1\over r_{jk}^2} 
{\Bigr ]}{\biggr)}.
\label{ddd}
\end{equation}
The integration over ${\bf r}_j$ is also straightforward and one finds,
\begin{equation}
2\pi \tau d\tau{\Bigr [}A-{2\pi \tau^2\beta^2 p^2\over 4} 
\sum_{k\not=l} p_k p_l \ln {r_{kl}\over\tau}{\Bigr ]}.
\label{djd}
\end{equation}
We can now write the second term of the RHS of Eq.~(\ref{aab}): 
\begin{eqnarray}
\sum_{N=1}^\infty &&y^{2N}{1\over (N!)^2}{1\over 2}\sum_{i\not=j}
\int_{D_{1}'} d^2r_1 ...\int_{D_{i-1}'} d^2r_{i-1}
\int_{D_{i+1}'} d^2r_{i+1} ...\int_{D_{j-1}'} d^2r_{j-1}\int_{D_{j+1}'} d^2r_{j+1} ...
\nonumber \\
&&\times
\int_{D_{2N}'} d^2r_{2N}\exp{\Bigr [} {\beta\over 2}\sum_{k\not= l;k,l\not= 
i,j}^{2N-2}p_k p_l \ln {r_{kl}\over \tau}
{\Bigr ]} {\Bigl (}2\pi \tau d\tau{\Bigr [}A-{2\pi \tau^2\beta^2 p^2\over 4} 
\sum_{k\not=l} p_k p_l \ln {r_{kl}\over\tau}{\Bigr ]}{\Bigr )}.
\label{eee}
\end{eqnarray}
After a change of variables in Eq.~(\ref{eee}) ($N'=N-1$), this equation can be 
added to the first term of the right hand side of Eq.~(\ref{aab}):
\begin{eqnarray}
Z=\sum_N y^{2N}&&{1\over (N!)^2}
\int_{D_{1}'} d^2r_1...
\int_{D_{2N}'} d^2r_{2N}\nonumber \\
&&{\Bigr [}1+{2\pi y^2 } \tau d\tau{\Bigr (} A-{2\pi 
\tau^2\beta^2 p^2\over 4} \sum_{k\not=l} p_k p_l
\ln {r_{kl}\over\tau}  {\Bigr )}{\Bigr ]}
\exp{\Bigr [} -{\beta\over 2}\sum_{i\not= j}p_i p_j V(r_{ij}){\Bigr ]}. 
\end{eqnarray}
Exponentiating the terms in the square bracket, 
\begin{eqnarray}
Z=\exp[{2\pi y^2 }&& \tau d\tau A]\sum_N y^{2N}{1\over (N!)^2}
\int_{D_{1}'} d^2r_1...\nonumber \\
&&\times 
\int_{D_{2N}'} d^2r_{2N}\exp{\Bigr \{} -{\beta\over 2}\sum_{i\not= j}p_i p_j {\Bigr [}-
\Bigl(1-(2\pi y\tau^2)^2{\beta p^2\over 2}{ d\tau\over\tau}\Bigr)\ln 
{r_{ij}\over\tau}{\Bigr ]} {\Bigr \}}. 
\label{fff}
\end{eqnarray}
Finally, we can rescale the lengths in Eq.~(\ref{fff}) so that the limits in the 
integral are the same as those of the original partition function using 
$r'=r/(1+d\tau/\tau)$,
\begin{eqnarray}
Z=\exp[&&{2\pi y^2 } \tau d\tau A] \sum_N \Bigl(y\Bigl [1+d\tau/\tau(2-\beta 
p^2/2)\Bigr]\Bigr)^{2N}{1\over (N!)^2}\nonumber \\
&&\int_{D_{1}} d^2r_1...
\int_{D_{2N}} d^2r_{2N}\exp{\Bigr \{} -{\beta\over 2}\sum_{i\not= j}p_i p_j  {\Bigr [}-
\Bigl(1-(2\pi y\tau^2)^2{\beta p^2\over 2}{ d\tau\over\tau}\Bigr)\ln 
{r_{ij}\over\tau}{\Bigr ]} {\Bigr \}}. 
\end{eqnarray}
One can see that this expression has the same form as the original except that the 
parameters have been renormalized:
\begin{equation}
(\beta p^2)'=\beta p^2(1-(2\pi y\tau^2)^2\beta p^2d\tau/2\tau),
\label{raa}
\end{equation}
\begin{equation}
y'=y(1+(2-\beta p^2
/2) d\tau/\tau)
\label{rbb}
\end{equation}
These equations can be put in differential form,
\begin{equation}
{d(\beta p^2)\over (\beta p^2)^2}=-d(\beta p^2)^{-1}=-{(2\pi y\tau^2)^2d\tau\over 2\tau},
\end{equation}
\begin{equation}
dy={y\beta p^2\over 2}({4\over\beta p^2} -1) d\tau/\tau.
\end{equation}
and the definitions $\epsilon=\ln(\tau/\xi_0)$, $y'=2\pi y\tau^2$, and $K=\beta 
p^2$ can be made yielding,
\begin{equation}
dK/d\epsilon= -y^2K^2/2,
\label{rra}
\end{equation}
\begin{equation}
dy/d\epsilon=y(4-K)/2,
\label{rrb}
\end{equation}
where we have dropped the primes. These equations are frequently written in terms of 
$x=4/K -1$: 
\begin{equation}
dx/d\epsilon=2 y^2,
\label{rrx}
\end{equation}
\begin{equation}
dy/d\epsilon=2yx,
\label{rry}
\end{equation}
where we have linearized the equations in $x$. These equations can be integrated 
to find $x(\epsilon)$ (or $K(\epsilon)$) and $y(\epsilon)$. 

\subsection{Analysis of the Recursion Relations}
\label{sec:twoc}

The recursion relations Eqs.~(\ref{rrx})-(\ref{rry}) have a fixed line at $y=0$ 
along which the parameters of the system do not change. But it is the fixed point 
[$x=0$,$y=0$] that is of more interest to us as we shall see. How these 
parameters change as one goes through the RG iterations are represented as 
``flows'' in phase space. By inspecting Eq.~(\ref{rrx}), one sees that $x$ can only 
increase under a RG iteration but that the increments can become infinitesimally 
small as $y$ goes to zero. Eq.~(\ref{rry}) on the other hand tells us that $y$ 
increases when $x$ is positive and decreases when $x$ is negative. This behavior 
is reflected in Fig.~\ref{rgflow}a where the RG flows are plotted in $x$-$y$ space 
for various initial values of $x$ and $y$. One can see that the lines end on the 
$y$ axis, tend toward [$x=\infty$,$y=\infty$], or, in one case, flow to 
[$x=0$,$y=0$]. This reflects the tendency of the RG iterations to drive the system 
to its ``pure'' limits: the low-temperature limit of no vortices ($y=0$) or the high-
temperature limit of a high vortex density. The last case where the flows go to the 
fixed point represents a sepatrix or, in RG parlance, a critical surface. This 
surface is associated with the critical temperature (where the system is scale-
invariant) since it does not flow to either the low or high temperature limits. 
Plotted in Fig.~\ref{rgflow}b are the RG flows for Eqs.~(\ref{rra})-(\ref{rrb}) 
with initial values of $K$ and $y$ corresponding to those used in 
Fig.~\ref{rgflow}a. The same behavior is reflected.

The first integral of Eqs.~(\ref{rrx})-(\ref{rry}) is 
\begin{equation}
y^2-x^2=c,
\label{firstI}
\end{equation}
where $c$ is a constant. Eq.~(\ref{firstI}) is the equation for a hyperbola and the 
flows of Fig.~\ref{rgflow} represent a family of hyperbolae corresponding to 
different values of $c$. Based on our discussion in the above paragraph, it is 
natural to make the assumption,
\begin{equation}
c\propto (T-T_{KTB})/T_{KTB}.
\label{TKTB}
\end{equation}
For $c>0$, the system is at a temperature above the critical one and the flows go 
to the high-temperature limit. For $c=0$, $T=T_{KTB}$ and the RG flow goes 
to the origin. For $c<0$, one is in the low-temperature limit.

With Eq.~(\ref{firstI}) and Eq.~(\ref{TKTB}), one can derive the temperature 
dependence for the vortex correlation length. Substituting Eq.~(\ref{firstI}) into 
Eq.~(\ref{rrx}), one has $dx/d\epsilon=2 (x^2+c)$. Integrating, one gets
\begin{equation}
{1\over\sqrt{c}} \arctan {x\over\sqrt{c}}\Biggr |_{x_i}^{x_{max}} = 2\epsilon,
\label{int}
\end{equation}
where $x_i$ is the bare or initial value of $x$ and $x_{max}$ is the final value. 
To be more precise, one integrates the recursion relations until there are no longer 
valid which is when the vortex density becomes large (i.e., when $y$ becomes of 
$O(1)$). The value of $\epsilon$ at that point is $\epsilon_{max}$. Because the 
correlation length $\xi$ is dominating the critical behavior and because the only 
length scale in Eq.~(\ref{int}) is $\tau=\xi_0\exp[\epsilon_{max}]$, one must 
make the association $\xi=\tau$. Then, for small $c$, one finds
\begin{equation}
\xi(T)\propto\exp[\sqrt{b/(T-T_{KTB})}].
\label{xiT}
\end{equation}
This result is valid for $T>T_{KTB}$ and for $T<T_{KTB}$ one can show that 
the correlation length is infinite (Kosterlitz 1974), which agrees with the result 
mentioned earlier that the susceptibility is infinite at small temperatures. This 
unique temperature dependence of the correlation length is a distinguishing 
property of the KTB transition. Typically, one has a power law divergence of the 
correlation length. The correlation length can be thought of as a screening length 
for the Coulombic interaction. For separations less than it, the interaction is 
logarithmic but approaches a constant at larger separations due to the screening 
of free vortices. It can also be interpreted as the average distance between free 
vortices.

Further insights into the KTB transition can be gained by looking at the length 
scale dependence of $K$ and $y$ as derived from Eqs.~(\ref{rra})-(\ref{rrb}). In 
Fig.~\ref{fig2}, we have plotted $K(\epsilon)$ and $y(\epsilon)$ for the same 
initial values of $K$ and $y$ as those used in Fig.~\ref{rgflow}b. From the 
length scale dependence of $y(\epsilon)$, one can see that there are two types of 
behavior as in Fig.~\ref{rgflow}b, one corresponding to $T>T_{KTB}$ and the 
other to $T<T_{KTB}$. In the latter, $y(\epsilon)$ falls monotonically to zero 
meaning that the density of vortices with a separation $R=\xi_0\ln\epsilon$ 
becomes negligible at large separations. Above $T_{KTB}$, this quantity 
initially decreases meaning again that the density of vortices falls with increasing 
separation, but then starts to increase. This increase marks the onset of unbinding 
of vortex pairs at the larger separations and the spontaneous creation of free 
vortices. This behavior is reflected in $K(\epsilon)$ depicted in Fig.~2b. For 
temperatures less than $T_{KTB}$, $K$ decreases due to the screening of vortex 
pairs but saturates at a finite value. The length scale at which it saturates 
corresponds to the length beyond which vortex pairs with the corresponding 
separation are negligible. The effective interaction is proportional to 
$K(R)\ln(R/\xi_0$ which remains largely logarithmic. This is in contrast to the 
$T>T_{KTB}$ case where one can see that $K$ does not saturate but decreases 
to zero. This is because vortices are present at all length scales. In fact, it is not 
just $K$ that changes but also the logarithmic dependence of the interaction 
energy due to free vortices. For further information, see the discussion by 
Minnhagen (1987).

\subsection{Observing the KTB transition in superconducting films}

An intriguing consequence of the KTB temperature dependence is that there is no 
singularity in the free energy of this system. Using a scaling analysis, one finds 
that the singular part of the free energy density scales as
\begin{equation}
f\propto\xi^{-2}
\label{fe}
\end{equation}
which is analytic as $T\longrightarrow T_{KTB}$. The specific heat and the 
magnetization, both thermodynamic derivatives of the free energy, are also 
analytic at the critical temperature leaving detection of this transition difficult. 

In superconducting films, it is primarily through electronic transport 
measurements that KTB critical behavior is verified. Here we will briefly sketch 
the arguments leading to the current-voltage characteristics and the temperature 
dependence of the resistance $R(T)$. In the presence of a current, a force (whose 
magnitude is linear with current and whose sign is opposite for vortices and anti-
vortices,) is exerted on a vortex in a direction perpendicular to the current. Free 
vortices can therefore dissipate energy. The resistance in the limit of zero current 
should be proportional to the density of free vortices which in turn is inversely 
proportional to the square of the correlation length:
\begin{equation}
R(T)\propto n_F\propto\xi^{-2}.
\label{RT}
\end{equation}
Minnhagen (1987) has generalized this formula to include the effects of the 
underlying superfluid to obtain $R(T)\propto\exp[\sqrt{b(T_{c0}-T)/(T-
T_{KTB})}]$ where $T_{c0}$ is the mean-field transition temperature. 

The effect of a current on a pair is to pull them apart with a force which is 
independent of their separation. Thus, the bare energy of a vortex pair in the 
presence of a current is 
\begin{equation}
E(R)=2E_c+p^2\log(R/\xi_0) -IR/\xi_0
\label{epair}
\end{equation}
where we have expressed the current in units of energy. At large $R$, the current 
repulsion will always dominate and $E(R)$ will have a peak at 
$R_c=\xi_0p^2/I$. Thermal energy will activate hops over this barrier providing 
a mechanism for vortex pair unbinding with a rate (Huberman, Myerson, and 
Doniach 1978)
\begin{equation}
\Gamma\propto\exp[-E_B/k_BT],
\label{rate}
\end{equation}
where $E_B=2E_c-p^2 +p^2\log(p^2/I)$ is the height of the barrier. The density 
of free vortices in this case can be found using the kinetic equation,
\begin{equation}
dn_F/dt=\Gamma-n_F^2,
\label{kinetic}
\end{equation}
where the second term on the right hand side account for vortex/anti-vortex 
pairing or annihilation. In the steady state, the RHS is zero so that $n_F=\Gamma^{1/2}$. 
This results in a non-linear $I$-$V$ relation:
\begin{equation}
V\propto I^{\alpha(T)+1},
\label{IV}
\end{equation}
where, neglecting renormalization effects, $\alpha(T)=p^2/2k_BT$. In the limit 
of weak currents, $\alpha(T)$ is decreasing linearly to a value, 
$\alpha(T_{KTB})=2$, as one increases the temperature to $T_{KTB}$ at which 
point it jumps discontinuously to zero. 

The $I$-$V$ signatures and the temperature dependence of $R(T)$ that 
characterize KTB behavior, Eqs.~(\ref{RT}) and (\ref{IV}), have been observed 
in superconducting films (Minnhagen 1987) providing strong evidence for the 
influence of vortex pair unbinding in these systems. They have also been observed 
in layered superconductors with a generalized form of Eq.~(\ref{IV}) (Jensen and 
Minnhagen 1991) indicating that vortices also play a major role there. We now 
proceed to discuss the layered system.

\section{Critical Behavior of Vortices in Layered Systems}   
\label{sec:body}

In this Section, the model and techniques of Section \ref{sec:model} are 
generalized to a weakly coupled layered system. The various properties of the 
critical behavior of vortices in layered systems deduced using the methods 
reviewed above will be emphasized but it should be noted that  many authors 
(Chattopadhyay and Shenoy 1994; Fischer 1993; Friesen 1995, Horovitz 1991; 
Korshunov; 1990) have made significant contributions to the field using other 
approaches. We will begin with the expected behavior of a layered system and the 
basic model used for such systems and discuss how the bare interactions of the 
vortices are modified in a layered system. The recursion relations for the 
interacting layered vortex gas derived from a RG analysis will then be given and 
an analysis of these recursion relations will be used to study the three-dimensional 
to two-dimensional crossover. Finally, the effect of a uniform electric current $I$ 
on the system and the $I$-$T$ phase diagram will be considered.

An intuitive understanding of the conventional wisdom (Leggett 1989) of the 
critical behavior of vortices in layered systems can be achieved by considering  
the correlation length which diverges at the critical temperature $T_c$. When the 
correlation length is smaller than the distance $d$ separating the layers, 2D 
behavior of the vortices is expected. When the correlation length becomes larger 
than the interlayer separation, the behavior should be 3D. Since the latter 
condition is met as the correlation length diverges near the transition, 3D 
behavior is expected in a small temperature window around $T_c$. Outside of 
this window, the layers are expected to be uncoupled. One can use 
$\xi_c=\lambda \exp[\sqrt{b/(T-T_{KTB})}]$ as an estimate for the correlation 
length in the $c$-direction (direction normal to the layers) where $\lambda$ is 
the ratio of the Josephson coupling energy to the in-plane coupling energy. Using 
the criterion $\xi_c=d$ for the dimensional crossover, one obtains for the size of 
the 3D critical region $\tau_{3D}$ a functional dependence, $\tau_{3D}\propto 
(\ln\lambda)^{-2}$. We will see below that this argument breaks down above 
$T_c$ where vortex fluctuations proliferate.

The most common model for a stack of weakly coupled superconducting layers is 
the Lawrence-Doniach model (Lawrence and Doniach, 1971) in which the 
coupling between the layers is taken to be the same as that between two 
superconductors separated by an insulator: Josephson coupling. The effect of the 
Josephson coupling on the vortex interactions is to strengthen the intralayer 
vortex interaction and to introduce interlayer vortex interactions. These 
interactions have been derived with the intralayer interaction $V(R,0)$ 
(Cataudella and Minnhagen 1990) being
\[ V(R,0)= \left \{ \begin{array}{ll}
\mbox{$-\ln(R/\tau)+(\lambda R^2/4\tau^2)\ln(\lambda R^2/\tau^2)$}, 
& \mbox{$\tau\ll R\ll R_\lambda$ } \\
\mbox{$-(\pi\sqrt{\lambda} R/\tau\sqrt{2})$}, &\mbox{$R\gg  R_\lambda$ }
\end{array}
\right.
\]
and the interlayer interaction $V(R,1)$ (Bulaevski, Meshkov, and Feinberg, 
1991) being
\[ V(R,1)\propto\left\{ \begin{array}{ll}
\mbox{$-\lambda R^2/\tau^2 \ln(\lambda R^2/\tau^2)$},  &\mbox{$R\ll R_\lambda$ }\\ 
\mbox{$\sqrt{\lambda} R/\tau$}, &\mbox{$R\gg R_\lambda $}
\end{array}
\right.
\]
where $R_\lambda=\xi_0/\sqrt\lambda$. For generality, one can introduce the 
effect of the current through the term,
\begin{equation}
H=\sum_i p_i {\bf r}_i\cdot {\bf J},
\end{equation}
where ${\bf r}_i$ is the in-plane Cartesian coordinate of the $i$th charge, 
$p_i=\pm\sqrt{d/2}\Phi_0/2\pi\Lambda$ is its charge, and J is related to the 
current $I$ by $J=\Phi_0 dI/p c a$ where $d$ is the distance between the 
layers, $\Phi_0$ is the superconducting flux quantum, $a$ is the 
characteristic cross-sectional area that relates the current to the 
current density $I/a$, $\Lambda$ is the London penetration depth, and $c$ is 
the speed of light. 

The partition function for the layered vortex gas (Pierson 1994; Pierson 1995b) in 
the presence of a current is,
\begin{eqnarray}
Z=\sum_N y^{2N}{1\over (N!)^2}
\sum_{l_1}\int_{D_{1}} d^2r_1 && \sum_{l_2}\int_{D_2} d^2r_2...\sum_{l_{2N}}
\int_{D_{2N}} d^2r_{2N} \nonumber \\
&& \times \exp{\Bigr [} -{\beta\over 2}\sum_{i\not= j}p_i p_j V(\mid 
{\bf r}_i-{\bf r}_j \mid, l_i-l_j)+\beta\sum_i p_i {\bf r}_i\cdot {\bf J}{\Bigr ]},
\end{eqnarray}
where $l_i$ is the index of the layer in which the $i$th particle is located.  
Interactions between vortices separated by more than one layer are neglected. A 
renormalization group analysis has been carried out on this partition function and 
the following recursion relations were found, (Pierson 1995a; Pierson 1995b) 
\begin{equation}
\label{rrra}
dx/d\epsilon=2y^2[1-\lambda/16+J^2/(1+x)],
\end{equation}
\begin{equation}
\label{rrrb}
dy/d\epsilon=2[x+(1/2)\lambda\ln\lambda]y/(1+x)+Jy/\sqrt{1+x},
\end{equation}
\begin{equation}
\label{rrrc}
d\lambda/d\epsilon=2\lambda\{1-4y^2[1+J^2/(1+x)]/(1+x)\},
\end{equation}
\begin{equation}
\label{rrrd}
dJ/d\epsilon=J,
\end{equation}
where a factor of $\sqrt\beta\tau$ is absorbed into $J$. Note that these recursion 
relations reduce to Eqs.~(\ref{rrx})-(\ref{rry}) for $J=0$ and $\lambda=0$. One 
can see that in Eqs.~(\ref{rrra})-(\ref{rrrd}) the current tends to counteract the 
effect of the Josephson coupling. This is expected because the current introduces a 
repulsion between vortices of opposite sign while the Josephson interaction 
strengthens the interaction. 

Considering first the zero current recursion relations, the critical behavior of 
vortices in a layered system can be addressed. This is done by examining the 
correlation length which is defined in the same way as in Section \ref{sec:twoc}: 
$\xi(T)=\xi_0\exp[\epsilon_{max}]$. The correlation length will diverge like the 
2D dependence sufficiently far above the transition temperature $T_c$ but will 
cross over to a power-law temperature dependence closer to the transition 
temperature. $T_c$ is easily derived from this quantity since it peaks at this 
temperature. By varying $\lambda$ one can determine the functional dependence 
of $T_c$ on the interlayer coupling to find (Pierson 1995a)
\begin{equation}
T_c\propto1/(\ln(\lambda))^2
\label{tc}
\end{equation}
in agreement with the results of Hikami and Tsuneto (1980). By comparing the
correlation length to the 2D correlation length, insights into the size of the 3D 
critical region can be found. One will find that sufficiently far from $T_c$ the 
two correlation lengths are nearly identical which is interpreted to mean that the 
layered system behaves two-dimensionally in that region. Closer to $T_c$ 
however there is a slow bifurcation which is taken to indicate crossover to the 3D 
region and the size of this region can be studied as a function of $\lambda$. It is 
convenient to divide the 3D temperature window into two parts, one above $T_c$: 
$\tau_{3D}^+$, and one below $T_c$: $\tau_{3D}^-$. It is found (Pierson 
1995a) that while $\tau_{3D}^-(\lambda)$ has the same logarithmic dependence 
on $\lambda$ as previously found for $\tau_{3D}$, $\tau_{3D}^+$ 
differs markedly: $\tau_{3D}^+\propto \lambda^{1/4}.$ A power-law dependence on 
$\lambda$ was also found subsequently by Friesen (1995):  
$\tau_{3D}^+\propto \lambda^{3/4}.$  The anisotropy of the 3D behavior around 
$T_c$ is dramatic as evidenced not only by the different dependences on 
$\lambda$ but also by the magnitude difference above and below the transition 
temperature (Pierson 1995a). This anisotropic behavior is attributed to the 
importance of the vortex fluctuations above $T_c$ which screen out both the 
interlayer and intralayer coupling.

How the system crosses over the 3D to 2D behavior can also be addressed 
(Pierson 1995c) by studying Eq.~(\ref{rrrc}), the recursion relation for 
$\lambda$. Below $T_c$ where $y$ is small, $\lambda$ grows to the isotropic 
value. This means that as the size of the vortex pairs becomes larger, 3D effects 
become more important. Above $T_c$, $\lambda$ grows initially but then starts 
to decrease at larger length scales due to vortex screening represented in the term 
proportional to $y^2$. The length scale at which $\lambda(\epsilon)$ peaks is the 
interlayer screening length $l_{3D/2D}$ and it is found to get larger as one 
lowers the temperature towards $T_c$ presumably diverging there as the 
intralayer screening length (i.e. the correlation length) does. The implication of 
this for the nature of 3D to 2D crossover is that the layer decoupling occurs first 
at the largest length scales just above $T_c$ and then proceeds to smaller length 
scales as the temperature is increased. As a consequence, the interaction between 
two vortices in neighboring layers separated by a distance larger than 
$l_{3D/2D}$ will be strongly screened giving the effect of decoupled layers. The 
linear (or 3D-like) interaction between two vortices in the same layers separated 
by a distance larger than interlayer screening length will also be screened 
resulting in the same effect. The screening of 3D effects at large length scales has 
interesting effects on the effective dimensionality of vortices where the 3D effects 
are felt primarily at those same lengths (Pierson 1995c).

An analysis of the finite current recursion relations has also shown to be fruitful 
(Pierson 1995b). By comparing the correlation length to the zero current 
correlation length and the 2D correlation length, one finds that there are three 
characteristic currents. The smallest $I_c^1(T)$ is the current that marks the 
onset of resistance. It corresponds to the minimum current needed to overcome 
the Josephson attraction to allow vortex pair unbinding. The next characteristic 
current $I_c(T)$ is where the phase transition occurs and above which free 
vortices can be spontaneously created. The largest characteristic current 
$I_c^2(T)$ is the 3D/2D crossover current. The temperature dependencies of 
these characteristic currents have been examined and are found to be 
predominantly linear. While $I_c^1(T)$  is linear at larger currents, it crosses 
over to join $I_c(T)$ at small currents converging to the same point at $T_c$. 
These three currents map out a phase diagram in $I$-$T$ space. The properties of 
electronic transport properties in each region of the $I$-$T$ phase diagram are 
discussed by the author (Pierson 1995b).  It should be stressed that $I_c^1(T)$  
and $I_c^2(T)$ are crossover currents while $I_c(T)$ is a phase transition.

\section{Summary}   
\label{sec:summ}

In this paper, we have reviewed the critical behavior of the 2D $X$-$Y$ model 
universality class with an emphasis on the derivation of the recursion relations 
through a renormalization group analysis. We reviewed the derivation of the 
temperature dependence of the correlation length and the methods of verifying the 
transition in superconductors. Finally those methods were generalized to a layered 
superconductor system where the vortex interactions are modified due to the 
Josephson coupling between the layers. The power and usefulness of the real 
space RG method become readily apparent here as evidenced by the calculation of 
a number of important properties of the critical behavior including the 
dependence of the transition temperature and the size of the 3D critical region on 
the strength of the interlayer coupling.

\acknowledgements   
This work was supported by the Office of Naval Research.

\begin{figure}   
\protect\caption{(a) The RG flows for the recursion relations 
Eqs.~(\protect\ref{rrx})-(\protect\ref{rry}) for various initial values of $x$ and 
$y$  [$x_i=-0.5$, $y_i=0.44, 0.46, \cdots 0.56$]; and (b) The RG flows for 
Eqs.~(\protect\ref{rra})-(\protect\ref{rrb}) for the initial values: $K_i=8.0$, 
$y_i=0.56, 0.58, \cdots 0.68$. Because Eqs.~(\protect\ref{rrx})-(\protect\ref{rry}) 
have been linearized around $x=0$, the initial values in (a) are not the same as 
those in (b). In both figures, there are two classes of flows, one for which $y$ goes 
to zero and one where $y$ becomes large, separated by a curve corresponding to 
$T=T_{KTB}$.} 
\label{rgflow}   
\end{figure}   

\begin{figure}   
\protect\caption{The length scale dependence of (a) $y$ and (b) $K$ plotted 
versus $\epsilon=\ln(\tau/\xi_0)$. In the low temperature phase, $K(\epsilon)$ 
saturates to a finite value at a length corresponding to the separation beyond 
which the vortex pair density becomes negligible. Above $T_{KTB}$ where the 
number of vortices increases at larger lengths, $K(\epsilon)$ is renormalized 
towards zero.}
\label{fig2}   
\end{figure}   
   
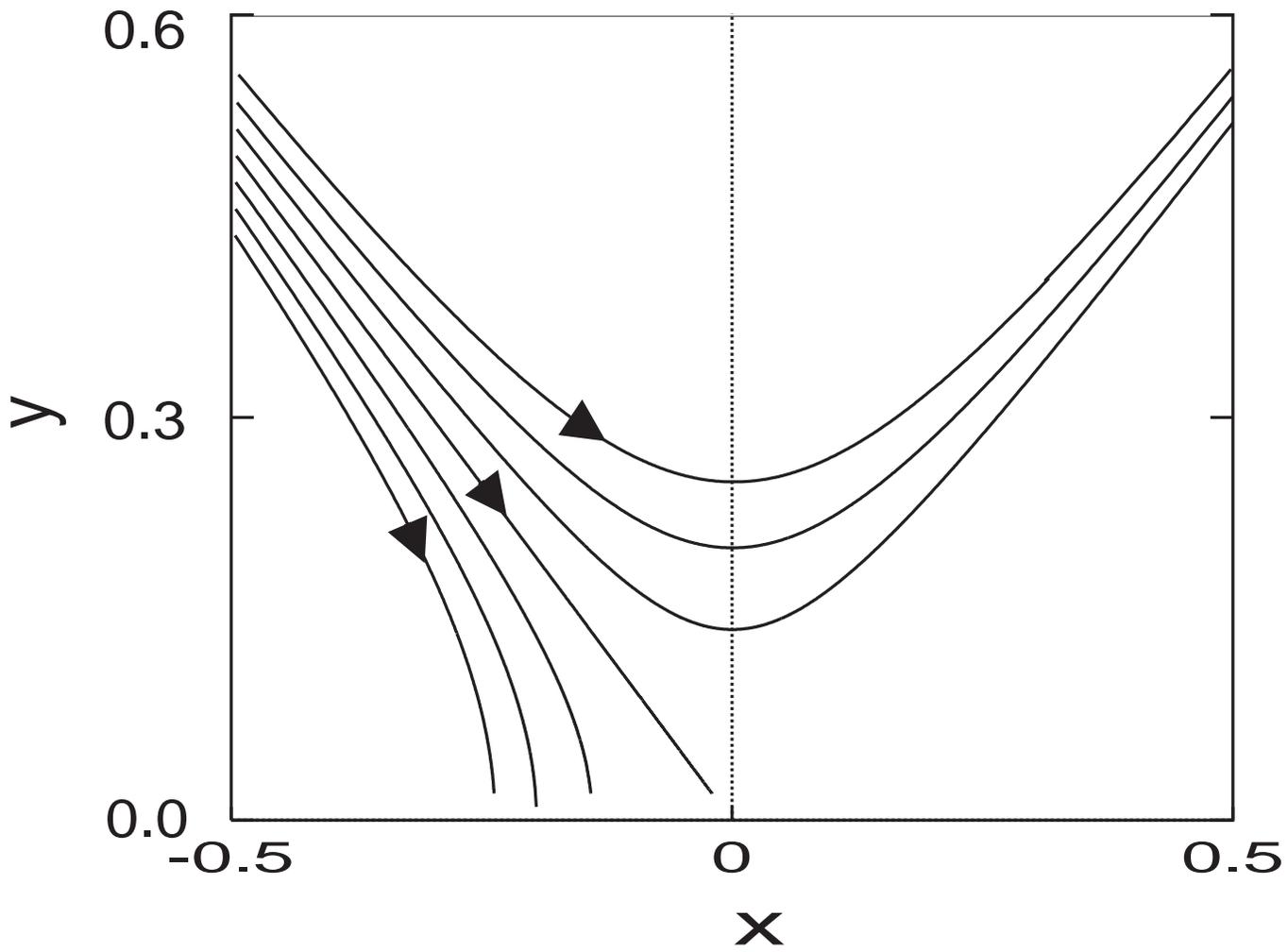
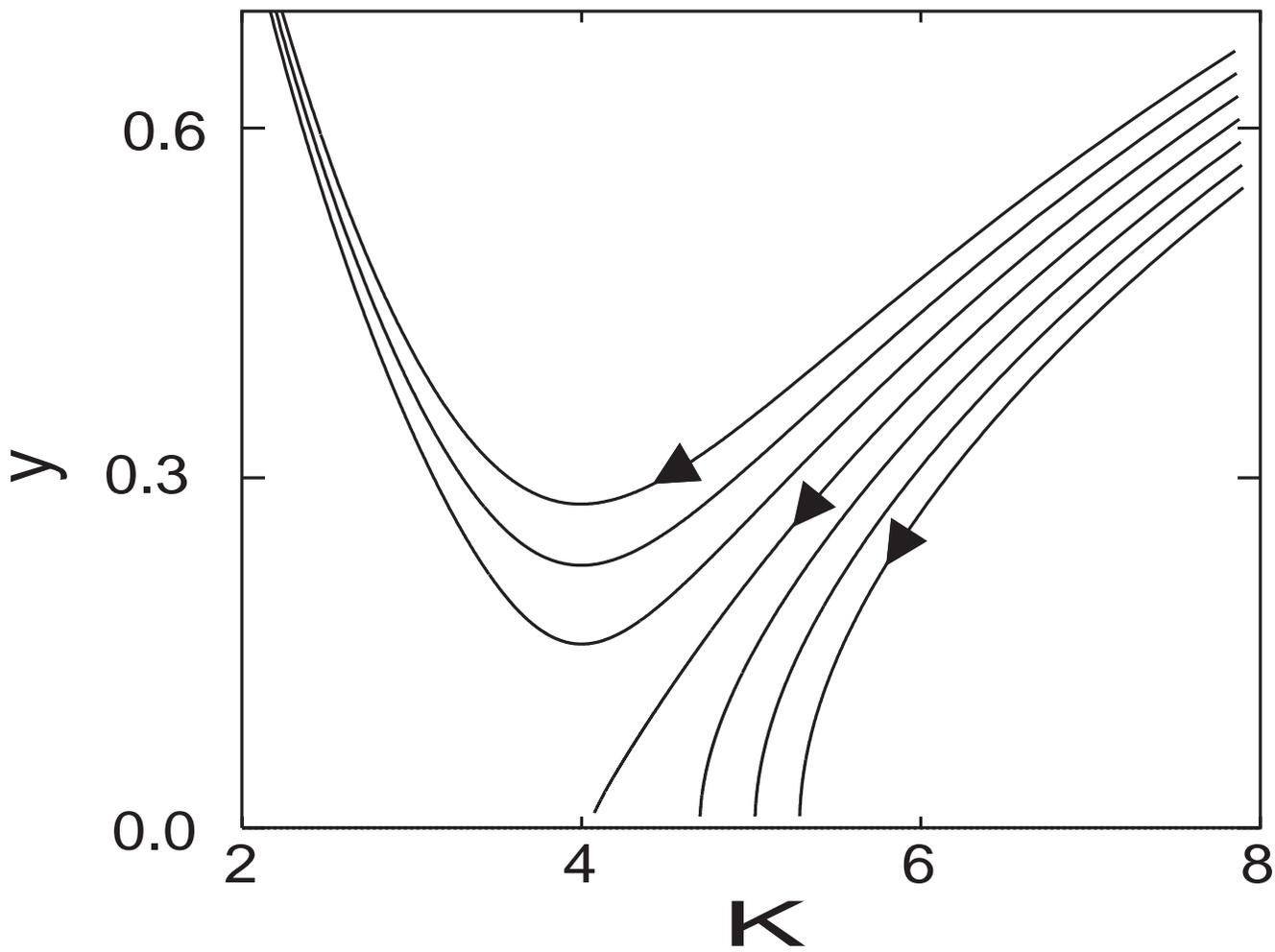
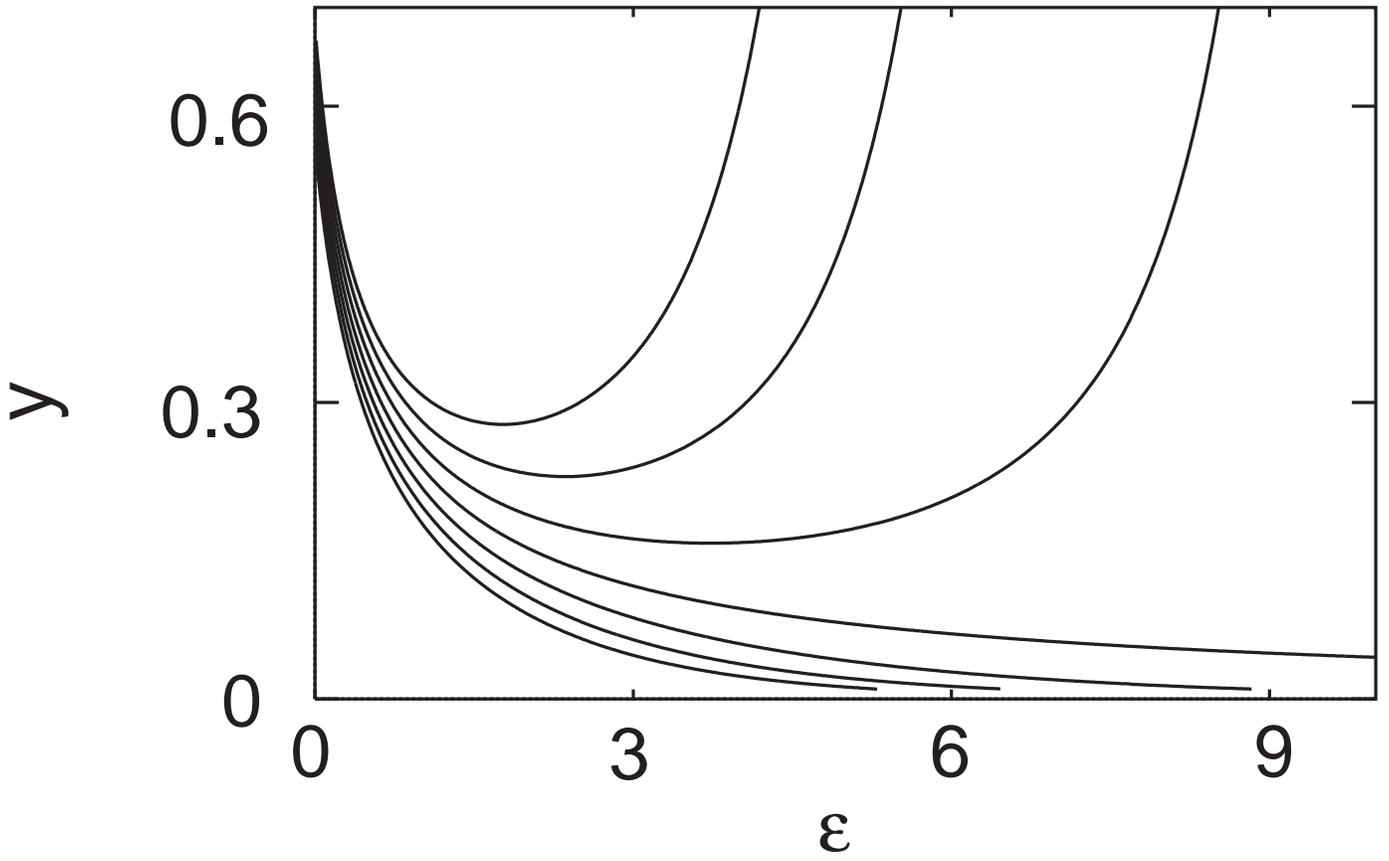
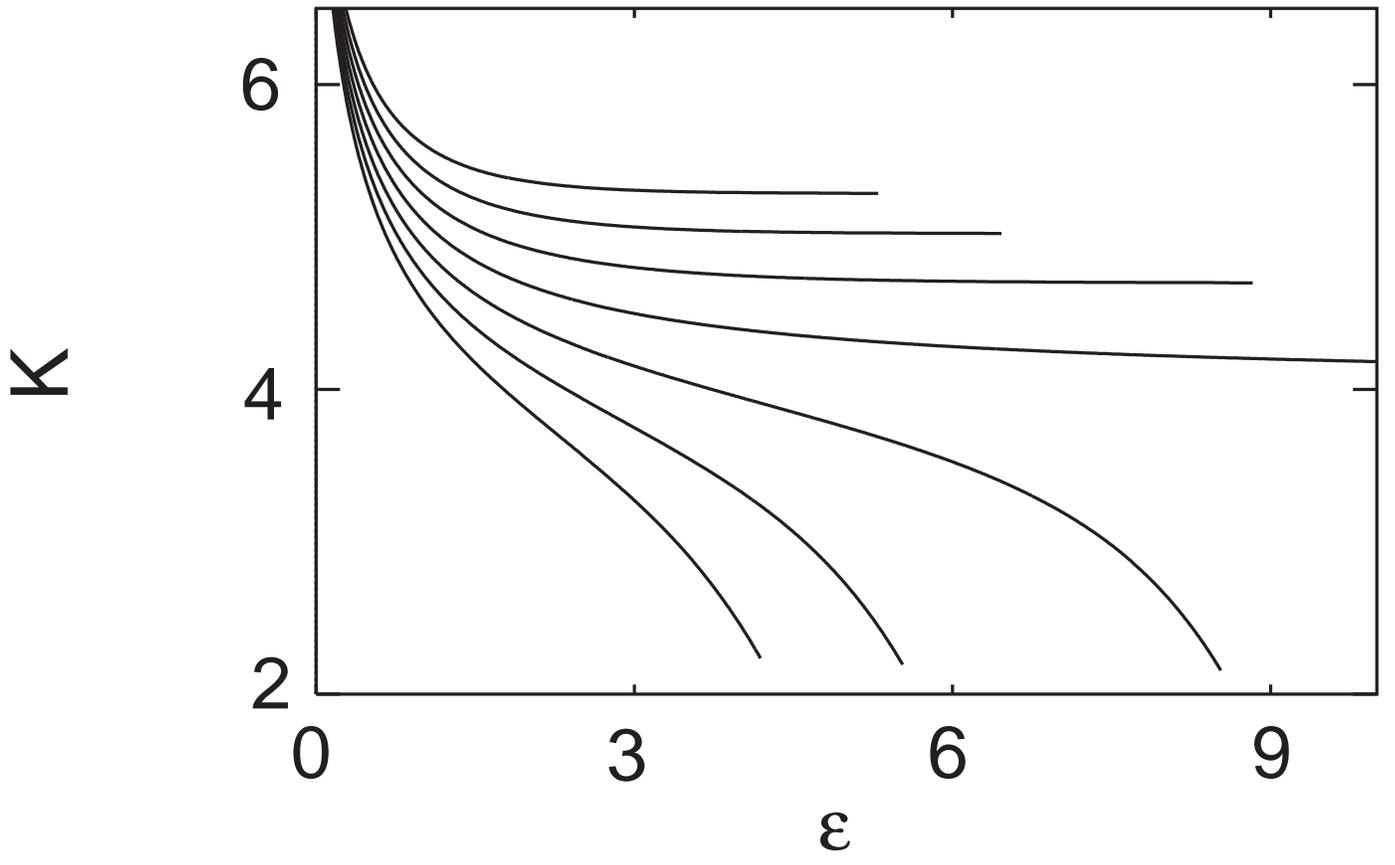
\end{document}